\documentclass[12pt]{article}

\usepackage{amssymb,amsfonts,amsthm}
\usepackage{amsmath}
\usepackage{bm}             
\usepackage[mathscr]{eucal}
\usepackage{cancel}
\usepackage{wasysym}

\def\be{\begin{equation}}
\def\ee{\end{equation}}
\def\bea{\begin{eqnarray}}
\def\eea{\end{eqnarray}}

\def\bna{\mbox{\boldmath $\nabla$}}

\def\bdelta{\mbox{\boldmath $\delta$}}

\def\hsp5{\hspace{5mm}}

\theoremstyle{remark}

\newcommand{\sfrac}[2]{{\textstyle{#1\over#2}}}

\title{\sc The general solution at large scale for second order perturbations in a
scalar field dominated universe}

\begin{document}

\author{ \\
{\Large\sc Claes Uggla}\thanks{Electronic address:
{\tt claes.uggla@kau.se}} \\[1ex]
Department of Physics, \\
Karlstad University, S-651 88 Karlstad, Sweden
\and \\
{\Large\sc John Wainwright}\thanks{Electronic address:
{\tt jwainwri@uwaterloo.ca}} \\[1ex]
Department of Applied Mathematics, \\
University of Waterloo,Waterloo, ON, N2L 3G1, Canada \\[2ex] }

\date{}
\maketitle
\bigskip

\begin{abstract}

In this paper we consider second order perturbations of a flat Friedmann-Lema\^{i}tre
universe whose stress-energy content is a single minimally coupled scalar field with an
arbitrary potential. We derive the general solution of the perturbed Einstein equations in
explicit form for this class of models when the perturbations are in the super-horizon
regime. As a by-product we obtain a new conserved quantity for long wavelength
perturbations of a single scalar field at second order.

\end{abstract}

\section{Introduction}

Cosmological perturbation theory plays a central role in confronting theories
of the early universe with observations. The increasing
accuracy of the observations, however,
has made it necessary to extend the theory from linear to
second order (\emph{i.e.\,}nonlinear)
perturbations, which presents various technical challenges.
This paper is a contribution
to this effort, focussing on long wavelength perturbations of
a flat Friedmann-Lema\^{i}tre (FL) universe whose
stress-energy content is a single
minimally coupled scalar field with an arbitrary potential.
For this class of models we derive the general solution of the
perturbed Einstein equations at second order
when the perturbations are in the super-horizon regime, including
both the growing and decaying modes.
This paper relies on three of our previous papers on
cosmological perturbation theory
which we shall refer to as UW1~\cite{uggwai19a}
(a unified and simplified formulation of change of gauge
formulas at second order), UW2~\cite{uggwai18}
(five ready-to-use systems of governing equations for second order perturbations)
and UW3~\cite{uggwai19b} (conserved quantities and the
general solution of the perturbed Einstein
equations for adiabatic long wavelength perturbations).

Our method is to apply the general solution in the total matter gauge given in
UW3~\cite{uggwai19b}  to the case of a scalar field and then transform to the uniform
curvature gauge.\footnote{Experience has shown that for these models the uniform
curvature gauge is the best choice to represent the perturbations of the scalar field.
See, for example, Hwang (1994)~\cite{hwa94b} (remarks in the Discussion), Liddle and
Lyth (2000)~\cite{lidlyt00} (page 93).}
The scalar field perturbations at first and second order are
algebraically related to the metric perturbations, and we show that in the uniform
curvature gauge, denoted by a subscript $_\mathrm{c}$, they have the following form:
\begin{equation}  \label{varphi_c.solution}
{}^{(1)}\!{\varphi}_{\mathrm c}\approx
(\varphi_0'){}^{(1)}\!C, \qquad{}^{(2)}\!{\varphi}_{\mathrm c}\approx
(\varphi_0'){}^{(2)}\!C -
(\varphi_0''){}^{(1)}\!C^2,
\end{equation}
where  $\varphi_0 = \varphi_0(N)$ is the background scalar field and $'$
denotes the derivative\footnote{We note that in cosmological perturbation
theory a $'$, in contrast to the present paper, is often used to
denote differentiation with respect to conformal time.}
with respect to $e$-fold time $N=\ln x = \ln(a/a_{init})$.
The background scalar field is determined
by the background Klein-Gordon equation:
\begin{equation} \label{KG_N}
\varphi_0'' +\sfrac12(6-(\varphi_0')^2)\left(\varphi_0' +
{V_{,\varphi}}/{V}\right)=0,
\end{equation}
where $V_{,\varphi}$ is the derivative of the
potential $V(\varphi_0)$ with respect to $\varphi_0$.
The arbitrary spatial functions ${}^{(1)}\!C$ and ${}^{(2)}\!C$
in equation~\eqref{varphi_c.solution} are
related to the comoving curvature perturbation
${}^{(r)}\!{\cal R}$, $r=1,2$, according to
\begin{equation}
{}^{(1)}\!C= {}^{(1)}\!{\cal R}, \qquad
{}^{(2)}\!C= {}^{(2)}\!{\cal R} + 2{}^{(1)}\!{\cal R}^2.
\end{equation}
Note that we have not imposed the slow-roll approximation
in obtaining this solution, and have not had to solve the perturbed
Klein-Gordon equation. As a by-product we obtain a new
conserved quantity for long wavelength perturbations of
a single scalar field at second order.

The outline of the paper is as follows.
In section~\ref{long.wavelength} we present the main results,
first the new conserved quantity at second order for a perturbed scalar field,
and then give explicit expressions for the scalar field perturbations
up to second order. In section~\ref{KG.equation} we derive a new form
of the perturbed Klein-Gordon equation which makes explicit the
existence of the conserved quantities at first and second order.
In appendix~\ref{scalar.field} we introduce the necessary background
material concerning scalar field perturbations.

\section{Long wavelength perturbations \label{long.wavelength}}

We consider second order scalar perturbations of a flat
FL universe with a single minimally coupled scalar field as matter.
Since we are going to specialize the general
framework developed in UW2~\cite{uggwai18} to this
situation we begin by briefly introducing the notation used in~\cite{uggwai18},
referring to that paper for further details. We
write the perturbed metric in the form\footnote{The scalar perturbations at
first order will generate vector and
tensor perturbations at second order, but we do not
give these perturbation variables since we will not consider these
modes in this paper.}
\begin{equation}  \label{pert.metric}
ds^2 = a^2\left(-(1+2\phi) d\eta^2 +  {\bf D}_i B\,d\eta dx^i +
(1-2\psi)\delta_{ij} dx^i dx^j \right),
\end{equation}
where $\eta$ is conformal time, the $x^i$ are Cartesian background coordinates
and ${\bf D}_i = \partial/\partial x^i$. The background geometry is described by the
scale factor $a$ which determines the conformal Hubble scalar ${\cal H}=a'/a$,
where in this specific situation $'$ denotes differentiation with respect to $\eta$.
By expanding the functions $\phi, B, \psi$ in a perturbation series\footnote{A
perturbation series for a variable $f$ is a Taylor series in a perturbation
parameter $\epsilon$, of the form
$f= f_0 +\epsilon\,{}^{(1)}\!f + \sfrac12 \epsilon^2\,{}^{(2)}\!f + \dots.$}
we obtain the following metric perturbations up to second order:
${}^{(r)}\!\phi, {\cal H}{}^{(r)}\!B, {}^{(r)}\!\psi$, $r=1,2,$
where the factor of ${\cal H}$ ensures that the $B$-perturbation is dimensionless, see
UW1~\cite{uggwai19a} and UW2~\cite{uggwai18}.
We use a perfect fluid stress-energy tensor~\eqref{pf} to describe
the matter-energy content, with the matter perturbations described by the variables
${}^{(r)}\!\bdelta,\, {\cal H}{}^{(r)}\!V, {}^{(r)}\!\Gamma$, $r=1,2,$
where ${}^{(r)}\!\bdelta = {}^{(r)}\!\rho/(\rho_0+p_0)$ is the density
perturbation, ${}^{(r)}\!V$ is the scalar velocity perturbation,
defined by writing\footnote{UW2~\cite{uggwai18}, section IIC. We note
that $V$ is the customary notation for the potential of a
scalar field, which we will also use in this paper. The context will eliminate
possible confusion. }
 $au_i={\bf D}_iV$, and
${}^{(r)}\!\Gamma$ is the non-adiabatic pressure perturbation.

The scalar field will be denoted by $\varphi$ with background field $\varphi_0$
and perturbations ${}^{(r)}\!\varphi$, $r=1,2$, and the potential
will be denoted by $V(\varphi_0)$. As is well known,
the stress-energy tensor of a minimally coupled scalar field~\eqref{sfT_ab}
in a cosmological setting can be written in the form of a perfect fluid,
which means that we can apply the above framework.
Using the relation between the two stress-energy tensors
the matter perturbations can be expressed in terms of
the scalar field perturbations and the metric perturbations.
We give the technical details in Appendix~\ref{app:scalar}.
In this paper at the outset
we need the relation between ${\cal H}{}^{(r)}\!V$
and ${}^{(r)}\!\varphi$ given by equations~\eqref{scalar1.3} and~\eqref{scalar2.3}
and the simplified expressions for ${}^{(r)}\!\Gamma$
given by~\eqref{gamma_sf}.

In this section, since we are considering long wavelength
perturbations, we will rely heavily on the results of UW3~\cite{uggwai19b}.

\subsection{A new conserved quantity\label{cons.quantity}}

When analyzing perturbed inflationary universes
two useful and complementary sets of variables are
the gauge invariants ${}^{(r)}\!{\varphi}_{\mathrm c}$, $r=1,2$,
the perturbations of the scalar field in the uniform curvature gauge
which are sometimes referred to as the \emph{Sasaki-Mukhanov
variables}\footnote{See for example Malik (2005)~\cite{mal05}, equations (3.14)
and (3.21).} and the gauge invariants ${}^{(r)}\!{\psi}_{sc},\,r=1,2$,
the curvature perturbations in the uniform field gauge.\footnote{See for
example, Maldacena (2003)~\cite{mal03}, section 3, in which both
descriptions are used and compared.}
Our analysis in this section will rely to a large extent on these variables.

At the outset we note that we will primarily use  $e$-fold time $N$
as the time variable. We will write $\partial_N f= \partial f/\partial N$
for brevity, and when $f$ is a background quantity we will use a $'$
as in the introduction,
for example $\partial_N \varphi_0\equiv\varphi_0'$.
We will also use the factor $l$, given by
\begin{equation} \label{def_lambda}
l :=  -(\varphi_0')^{-1},
\end{equation}
as a shorthand notation to represent the frequent divisions by $\varphi_0'$
that occur in the equations involved in
the study of perturbations of scalar fields.

Our first goal is to derive a general relation between
the scalar field perturbations ${}^{(r)}\!{\varphi},\,r=1,2,$
and the metric perturbations, specifically the curvature perturbations
${}^{(r)}\!\psi,\,r=1,2$. We begin by performing a change of gauge from the uniform
field gauge, defined by ${}^{(r)}\!{\varphi}=0,\,r=1,2$, to an arbitrary gauge
using the formula (42e) in UW1~\cite{uggwai19a} with $\Box=\psi$:
\begin{subequations} \label{gauge.change}
\begin{align}
{}^{(1)}\!{\psi}_{\mathrm sc} &= {}^{(1)}\!{\psi} - l{}^{(1)}\!{\varphi},\\
{}^{(2)}\!\hat{\psi}_{\mathrm sc} &= {}^{(2)}\!\hat{\psi} -
l{}^{(2)}\!\hat{\varphi} +
2l{}^{(1)}\!{\varphi}\,\partial_N {}^{(1)}\!\psi_{\mathrm sc}  -
{\mathbb D}_2({}^{(1)}\!B_{\mathrm sc}) + {\mathbb D}_2({}^{(1)}\!B),\label{psi-varphi2}
\end{align}
\end{subequations}
where ${\mathbb D}_2$, the so-called Newtonian spatial operator,\footnote{See
UW1~\cite{uggwai19a}, Appendix B. The specific form
of ${\mathbb D}_2(\bullet)$ does not concern us here.} is a spatial differential operator
of order $2$ in ${\bf D}_i$ and hence negligible in the super-horizon regime.
The hatted variables are given by (see UW1~\cite{uggwai19a}, equations (37)):
\begin{subequations}
\begin{align}
{}^{(2)}\!{\hat \psi} :&={}^{(2)}\! \psi + 2\,{}^{(1)}\! \psi^2,  \\
l {}^{(2)}\!{\hat{\varphi}} :&=l {}^{(2)}\!{\varphi}  +
\sfrac32(w_{\varphi} -c_{\varphi}^2)(l{}^{(1)}\!{\varphi})^2,
\label{hat.varphi}
\end{align}
\end{subequations}
where $w_{\varphi}-c_{\varphi}^2$ is related to $\varphi_0$
and its derivatives by equation~\eqref{w-c^2} in appendix~\ref{background}.

It is helpful to use the fact that the uniform field gauge
is equivalent to the total matter gauge,
defined by ${}^{(r)}\!V=0$, $r=1,2$,
since equations~\eqref{scalar1.3} and~\eqref{scalar2.3}
show that ${}^{(r)}\!\varphi=0 \Leftrightarrow {}^{(r)}\!V=0$, $r=1,2$.
This means that the various gauge invariants in the
two gauges are equal. For example, for the curvature perturbation we have
\begin{equation}  \label{psi_sc=psi_v}
{}^{(r)}\!\psi_{\mathrm {sc}} = {}^{(r)}\!\psi_{\mathrm v},\qquad r=1,2.
\end{equation}
By solving for the $\varphi$-perturbation in equations~\eqref{gauge.change}
 and using equation~\eqref{psi_sc=psi_v} we obtain
\begin{subequations}  \label{varphi-psi}
\begin{align}
l{}^{(1)}\!{\varphi} &= {}^{(1)}\!{\psi} - {}^{(1)}\!{\psi}_{\mathrm v},\label{varphi.1}\\
l{}^{(2)}\!\hat{\varphi} &= {}^{(2)}\!\hat{\psi} -
{}^{(2)}\!\hat{\psi}_{\mathrm v} +
2({}^{(1)}\!{\psi} - {}^{(1)}\!{\psi}_{\mathrm v})\partial_N {}^{(1)}\!\psi_{\mathrm v}  -
{\mathbb D}_2({}^{(1)}\!B_{\mathrm v}) + {\mathbb D}_2({}^{(1)}\!B),\label{varphi.2}
\end{align}
\end{subequations}
which determine the scalar field perturbations in any
gauge in terms of the metric perturbations.

At this stage we need two general properties of
long wavelength perturbations:\footnote{These results are part of the folklore
of perturbation theory, but derivations at second order are not easy to find. We
have given simple derivations in UW3~\cite{uggwai19b}, equations (20) and (26a).}
\begin{itemize}
\item[i)] The density perturbations in the total matter gauge satisfy,
\begin{equation} \label{delta_v.zero.new}
{}^{(1)}\! {\bdelta}_{\mathrm v}\approx 0, \qquad {}^{(2)}\! {\bdelta}_{\mathrm v}\approx 0.
\end{equation}
\item[ii)] If the perturbations are adiabatic the curvature
perturbations in the total matter gauge satisfy,
\begin{equation} \label{deriv.psi_v.new}
\partial_N{}^{(1)}\! \psi_{\mathrm v} \approx 0, \qquad
 \partial_N{}^{(2)}\! \psi_{\mathrm v} \approx 0.
\end{equation}
\end{itemize}
In addition we need to determine the non-adiabatic pressure perturbations
for a scalar field. In Appendix~\ref{app:scalar} we have shown
that the ${}^{(r)}\!\Gamma, r=1,2$, depend linearly on
${}^{(r)}\!\bdelta_{\mathrm v}, r=1,2$, with source terms at second order
depending  on ${}^{(1)}\!\bdelta_{\mathrm v}$, as in equation~\eqref{gamma_sf}.
It thus follows from
 equation~\eqref{delta_v.zero.new} that
\begin{equation} \label{gamma_0.new}
{}^{(1)}\!\Gamma\approx 0, \qquad
{}^{(2)}\!\Gamma\approx 0,
\end{equation}
{\emph i.e.} long wavelength scalar field perturbations are adiabatic.\footnote
{This result has been given by Vernizzi (2005)~\cite{ver05}.}
Thus~\eqref{deriv.psi_v.new} holds,
which implies that for long wavelength perturbations
equations~\eqref{varphi-psi} reduce to
\begin{equation} \label{varphi-psi.long}
l{}^{(1)}\!{\varphi} = {}^{(1)}\!{\psi} - {}^{(1)}\!{\psi}_{\mathrm v}, \qquad
l{}^{(2)}\!\hat{\varphi} \approx {}^{(2)}\!\hat{\psi} -
{}^{(2)}\!\hat{\psi}_{\mathrm v}.
\end{equation}
In other words, when using the hatted variables the first order
relation generalizes to second order.

We now choose the arbitrary gauge in these equations to
be the uniform curvature gauge, which gives
\begin{equation} \label{varphi-psi.long.uc}
l{}^{(1)}\!{\varphi}_{\mathrm c} =  - {}^{(1)}\!{\psi}_{\mathrm v}, \qquad
l{}^{(2)}\!\hat{\varphi}_{\mathrm c} \approx  -
{}^{(2)}\!\hat{\psi}_{\mathrm v}.
\end{equation}
Since~\eqref{deriv.psi_v.new} holds the first equation
in~\eqref{varphi-psi.long.uc}
gives the known result\footnote{Sasaki
(1986)~\cite{sas86} (see equation (2.33))  introduced
the quantity ${}^{(1)}\!{\psi}_{\mathrm p} - l{}^{(1)}\!{\varphi}_{\mathrm p}$
in our notation, and stated that it is constant on large scales provided that the entropy and spatial
anisotropy perturbations are negligible.}
that at first order $l{}^{(1)}\!{\varphi}_{\mathrm c}$ is a
conserved quantity while
the second equation
gives the new result  that at second order
\begin{equation} \label{hat.varphi.new}
l{}^{(2)}\!\hat{\varphi}_{\mathrm c}:=
l {}^{(2)}\!{\varphi}_{\mathrm c}  +
\sfrac32(w_{\varphi} -c_{\varphi}^2)(l{}^{(1)}\!{\varphi}_{\mathrm c})^2,
\end{equation}
is a conserved quantity for long wavelength perturbations of a single field
inflationary universe. We note that the relation~\eqref{hat.varphi.new} is obtained
by choosing the uniform curvature gauge in~\eqref{hat.varphi}.

\subsection{Explicit form of the scalar field perturbations}

In a recent paper UW3~\cite{uggwai19b} we gave
the general solution of the perturbed Einstein
equations at second order for long wavelength
adiabatic perturbations of a FL universe, with stress-energy
tensor of the perfect fluid form~\eqref{pf}.
We solved the equations for the metric
perturbations in the total matter gauge, giving the general solution,
\emph {i.e.} including the decaying mode.
At first order the solution is
\begin{subequations} \label{tm_sh1}
\begin{equation} \label{comov_sh1.new}
{}^{(1)}\!\phi_{\mathrm v}\approx0, \qquad
{}^{(1)}\!\psi_{\mathrm v}\approx {}^{(1)}\!C,
\end{equation}
\begin{equation}
{\cal H}{}^{(1)}\!B_{\mathrm v} \approx (1-g(a)){}^{(1)}\!C
+({\cal H}/a^2) {}^{(1)}\!C_{*},
\label{comov_sh1_B.new}
\end{equation}
\end{subequations}
where the perturbation evolution function\footnote{We refer to UW3~\cite{uggwai19b}
for this name, and for its history and properties (see Appendix C in~\cite{uggwai19b}).}
$g(a)$ is defined by
\begin{equation} \label{def_g_simple}
g(a) = 1 - \frac{{\cal H}(a)}{a^2} \int_0^a \frac{\bar a}{{\cal H}(\bar a)}d{\bar a}.
\end{equation}
At second order the solution is:\footnote{The spatial differential
operator ${\mathbb D}_0$ is defined by
${\mathbb D}_0(C) := {\cal S}^{ij}({\bf D}_iC)({\bf D}_jC)$,
where the scalar mode extraction operator ${\cal S}^{ij}$
is defined by
${\cal S}^{ij} = \sfrac32({\bf D}^{-2})^2{\bf D}^{ij}$.
Here ${\bf D}^{-2}$ is the inverse spatial Laplacian and
${\bf D}_{ij} := {\bf D}_{(i}{\bf D}_{j)} - \sfrac13 \delta_{ij}{\bf D}^2$. Spatial
indices are raised with $\delta^{ij}$.}

\begin{subequations} \label{tm_sh2}
\begin{equation} \label{comov_sh2.new}
{}^{(2)}\!\phi_{\mathrm v}\approx0, \qquad
{}^{(2)}\!\hat{\psi}_{\mathrm v}\approx {}^{(2)}\!C,
\end{equation}
\begin{equation}
{\cal H}{}^{(2)}\!B_{\mathrm v} \approx\,(1-g(a))\left({}^{(2)}\!C -
2{\mathbb D}_0( {}^{(1)}\!C)\right)+({\cal H}/a^2) {}^{(2)}\!C_{*}. \label{B_v2.new}
\end{equation}
\end{subequations}
We identify the spatial functions ${}^{(1)}\!C(x^i)$ and
${}^{(2)}\!C(x^i)$ as the conserved
quantities at first and second order, while
${}^{(1)}\!C_{*}(x^i)$ and
${}^{(2)}\!C_{*}(x^i)$ describe the decaying mode.
If we apply this solution in the case of a scalar field the Hubble scalar ${\cal H}(a)$
is determined explicitly in terms of
the background scalar field and the scalar field
potential by equation~\eqref{H.ito.phi}, which
then determines the function $g(a)$ through~\eqref{def_g_simple}.

In the total matter gauge it follows from~\eqref{varphi-psi.long}
that the scalar field perturbations
are zero (${}^{(r)}\!\varphi_{\mathrm v}=0,\,r=1,2$). In effect, in this gauge
the perturbations of the scalar field are hidden in the metric perturbations.
On the other hand, in the uniform curvature gauge it follows
from~\eqref{varphi-psi.long.uc}  that the scalar field perturbations
are given by
\begin{equation} \label{phi_c.conserved}
l{}^{(1)}\!\varphi_{\mathrm c}=-{}^{(1)}\!\psi_{\mathrm v}\approx-{}^{(1)}\!C , \qquad
l{}^{(2)}\!\hat{\varphi}_{\mathrm c}\approx -{}^{(2)}\!\hat{\psi}_{\mathrm v}
\approx-{}^{(2)}\!C,
\end{equation}
the last step following from~\eqref{comov_sh1.new} and~\eqref{comov_sh2.new}.
To obtain an explicit expression for ${}^{(2)}\! {\varphi}_{\mathrm c}$ we
use the definition~\eqref{hat.varphi.new} of ${}^{(2)}\!\hat {\varphi}_{\mathrm c}$
and $l=-1/(\varphi_0')$ which yields:
\begin{equation} \label{varphi_2.temp.new}
{}^{(1)}\!{\varphi}_{\mathrm c}\approx
\varphi_0'{}^{(1)}\!C, \qquad
{}^{(2)}\!{\varphi}_{\mathrm c}\approx
\varphi_0'\left({}^{(2)}\!C + \sfrac32(w_{\varphi}-c_{\varphi}^2){}^{(1)}\!C^2\right),
\end{equation}
where $w_{\varphi}-c_{\varphi}^2$ is given
by equation~\eqref{w-c^2} in appendix~\ref{background}, which we also give here:
\begin{equation} \label{w-c^2.repeat}
w_{\varphi}-c_{\varphi}^2 =
\frac23\left(\frac{\varphi_0''}{\varphi_0'}\right).
\end{equation}
Substituting this result
into~\eqref{varphi_2.temp.new} yields the expression~\eqref{varphi_c.solution}
for ${}^{(2)}\!{\varphi}_{\mathrm c}$ in the introduction.
The scalar field $\varphi_0(N)$ is determined by the background
Klein-Gordon equation~\eqref{KG_N}.

We now calculate the scalar field perturbation in the
Poisson gauge by choosing the Poisson gauge in
equation~\eqref{varphi-psi.long} which yields:
\begin{equation} \label{varphi-psi.long.poisson}
l{}^{(1)}\!{\varphi}_{\mathrm p} =
{}^{(1)}\!{\psi}_{\mathrm p} - {}^{(1)}\!{\psi}_{\mathrm v}, \qquad
l{}^{(2)}\!\hat{\varphi}_{\mathrm p} \approx {}^{(2)}\!\hat{\psi}_{\mathrm p} -
{}^{(2)}\!\hat{\psi}_{\mathrm v}.
\end{equation}
In our previous paper UW3~\cite{uggwai19b} we used
a change of gauge formula to calculate
the curvature $\psi_{\mathrm p}$ determined by the
solution~\eqref{tm_sh1} and~\eqref{tm_sh2}.
The results are:
\begin{subequations}
\begin{align}
\psi_{\mathrm p}& \approx g C -({\cal H}/a^2) C_{*},\\
{}^{(2)}\! {\hat\psi}_{\mathrm p} &\approx \,g{}^{(2)}\!C +
(1-g)[\left((1+q)(1-g) - g\right) C^2 +
4g{\mathbb D}_0(C)].
\end{align}
\end{subequations}
where we note that for brevity have not included the decaying mode at
second order. We substitute these expressions into~\eqref{varphi-psi.long.poisson}
and use~\eqref{comov_sh1.new} and~\eqref{comov_sh2.new}.
The final expressions for the scalar field perturbations, using the
unhatted variable and excluding the decaying mode at
second order, are as follows:\footnote
{The decaying mode contribution to $l {}^{(2)}\!{\varphi_{\mathrm p}}$ has the
leading order term $-({\cal H}/a^2){}^{(2)}\!C_{*}$, where
${}^{(2)}\!C_{*}$ is another arbitrary spatial function, together with quadratic
source terms that are a linear combination of the following spatial functions,
$C_{*}^2, {\mathbb D}_0(C_{*}), CC_{*}, {\cal S}^i(C{\bf D}_iC_{*}),
{\cal S}^{ij}({\bf D}_iC{\bf D}_jC_{*})$, with time dependent coefficients.
The scalar mode extraction operator ${\cal S}^i$ is defined by
${\cal S}^i={\bf D}^{-2}{\bf D}^i.$}
\begin{subequations}
\begin{align}
l {}^{(1)}\!{\varphi_{\mathrm p}} &\approx \,-(1-g){}^{(1)}\!C -
({\cal H}/a^2){}^{(1)}\!C_{*},\\
l {}^{(2)}\!{\varphi_{\mathrm p}} &\approx
(1-g)\left[{-}^{(2)}\!C +
\left(\sfrac32 (1+c_{\varphi}^2)(1-g) -g\right) C^2  +4g\,{\mathbb D}_0(C)\right].
\end{align}
\end{subequations}
The linear solution is well known but is usually found by
solving the Bardeen equation or the perturbed Klein-Gordon equation.\footnote{See,
for example Mukhanov (2005)~\cite{muk05}, equation (8.68), or
Mukhanov (1985)~\cite{muk85}, equation (14),
 which reads
${}^{(1)}\!{\varphi}_{\mathrm p}=A\dot\varphi_0 a^{-1}\int a dt, $
where  $\dot \varphi_0=d\varphi_0/dt=-H/l.$ The expression
for $g(t)$ in UW3~\cite{uggwai19b} (see section 7 ) relates Mukhanov's result to ours.}
The second order solution is new.
Note that the decaying mode appears in the scalar field perturbations in the
Poisson gauge, in contrast to the situation in the uniform
curvature gauge in equation~\eqref{varphi_2.temp.new}.\footnote
{This property of the decaying mode at linear order has been pointed out previously, see for example Hwang (1994)~\cite{hwa94b}, table 1. It can also be inferred from
Brandenberger and Finelli (2001)~\cite{brafin01}.  }

We end this section with a comment on multiple scalar fields.
We note that perturbations of universes with multiple scalar fields
(see for example, Malik and Wands (2005)~\cite{malwan05}, for linear perturbations
and Malik (2005)~\cite{mal05} for second order perturbations) have been studied
using the uniform curvature gauge. However, since large scale perturbations of multiple scalar fields are not adiabatic, our explicit large scale solution for a single scalar field
cannot be generalized to multiple fields.

\section{The perturbed Klein-Gordon equation\label{KG.equation}}

Although the Klein-Gordon equation plays a central role
in governing scalar field perturbations,
we have obtained simple expressions for
these perturbations to second order on super-horizon scale
and have obtained conserved quantities,
using only some of the perturbed Einstein equations.
On inspecting the form of the perturbed Klein-Gordon equation at first and
second order as given in the literature\footnote{
See, for example,
Huston and Malik (2009)~\cite{husmal09}, equations
(2.13) and (2.15), in Fourier space (${\bf D}^2\rightarrow-k^2$). }
it is clear that our results are unexpected, since it is not obvious that
the equation admits conserved quantities or that  the explicit expressions
for ${}^{(1)}\!{\varphi}_{\mathrm c}$ and ${}^{(2)}\!{\varphi}_{\mathrm c},$
as given by equation~\eqref{varphi_c.solution}, are in fact solutions
of the equation on super-horizon scale.

There are two standard ways to derive the perturbed Klein-Gordon
equation. The first is to use the perturbed energy conservation equation
to obtain an expression for the second time
derivative of ${\varphi}_{\mathrm c}$ and then use
(some of) the perturbed Einstein equations to express the metric perturbations
that appear in this equation in terms of the scalar field perturbations. The second is to use
the variation of the Einstein action coupled to the scalar field. We refer to
Malik \emph{et al} (2008)~\cite{maletal08} for a comparison of the two approaches.

In this section we give a new way of deriving the
perturbed Klein-Gordon equation, leading to a simpler form of
the equation with $l=-1/(\varphi_0')$
acting as a scale factor for the perturbation, which
by inspection admits a conserved quantity at first and second order.
We begin with the perturbed Einstein equations
in the uniform curvature gauge as given in our earlier
paper UW2~\cite{uggwai18}, see section IVB.1.
These equations, which determine the
metric perturbations $\phi_\mathrm{c}$ and $B_\mathrm{c}$ at first order,
read
\begin{subequations}\label{ucg_gov1}
\begin{align}
\partial_N((1+q)^{-1} {}^{(1)}\!{\phi_\mathrm{c}}) &=
 -c_s^2(1+q)^{-1}{\cal H}^{-2}{\bf D}^2({\cal H}{}^{(1)}\!{B}_{\mathrm c})
 + {}^{(1)}\Gamma, \label{ucg_gov1.1} \\
\partial_N(a^2\,{}^{(1)}\!B_{\mathrm c} )&= -a^2 {\cal H}^ {-1}{}^{(1)}\!{\phi_\mathrm{c}}.  \label{ucg_gov1.2}
\end{align}
\end{subequations}
Here $q$ denotes the background deceleration parameter which is
defined by
\begin{equation}  \label{q.definition}
1+q=-H'/H,\,\Longleftrightarrow \,q={\cal H}'/{\cal H},
\end{equation}
where $H$ is the background Hubble scalar, ${\cal H}=aH$ and $'$ denotes
differentiation with respect to $N$.
The velocity perturbation and density perturbation are given by
\begin{subequations} \label{sc.field.relations}
\begin{align}
{\cal H}{}^{(1)}\!V_{\mathrm c} &=  -(1+q)^{-1} {}^{(1)}\!\phi_\mathrm{c},\\
{}^{(1)}\!{\bdelta}_{\mathrm v} &=  -
(1+q)^{-1}{\cal H}^{-2}{\bf D}^2 ({\cal H}{}^{(1)}\!{B}_{\mathrm c}).
\end{align}
In a universe with a single scalar field
equations~\eqref{scalar1.3}
and~\eqref{gamma_sf1}  in appendix~\ref{app:scalar} give
\begin{equation} \label{infl.uni}
{\cal H}{}^{(1)}\!V_{\mathrm c} = l{}^{(1)}\!\varphi_{\mathrm c}, \qquad
{}^{(1)}\!\Gamma=(1-c_s^2){}^{(1)}\!{\bdelta}_{\mathrm v},
\end{equation}
\end{subequations}
where $l=-1/(\varphi_0')$.
Using equations~\eqref{sc.field.relations}, which show that
\begin{equation} \label{phi.varphi}
(1+q)^{-1}{}^{(1)}\!{\phi_\mathrm{c}}= -l{}^{(1)}\!\varphi_{\mathrm c},
\end{equation}
we can write equations~\eqref{ucg_gov1} as a coupled system for
${\varphi}_{\mathrm c}$ and $B_{\mathrm c}$:
\begin{subequations}\label{ucg1_varphi}
\begin{align}
(1+q)\partial_N(l{}^{(1)}\!\varphi_{\mathrm c}) &=
{\cal H}^{-2}{\bf D}^2({\cal H}{}^{(1)}\!B_{\mathrm c}), \label{ucg1_varphi.1} \\
\partial_N(a^2\,{}^{(1)}\!B_{\mathrm c} )&=
a^2 {\cal H}^ {-1}(1+q)l{}^{(1)}\!\varphi_{\mathrm c}.  \label{ucg1_varphi.2}
\end{align}
\end{subequations}
We now eliminate
the metric perturbation ${}^{(1)}\!B_{\mathrm c}$  by
applying ${\bf D}^2$ to~\eqref{ucg1_varphi.2}
and substituting for ${\bf D}^2({\cal H}{B}_{\mathrm c})$ from~\eqref{ucg1_varphi.1}.
This leads to
\begin{equation} \label{DE.varphi.1}
\left(\partial_N^2+2\frac{h' }{h} \partial_N
- {\cal H}^{-2}{\bf D}^2\right)(l {}^{(1)}\!{\varphi}_{\mathrm c}) = 0,
\end{equation}
where $h=h(N)$ is a background scalar given by\footnote
{We use $h^2$ instead of $h$ in order to have a simple link
with the commonly used Mukhanov-Sasaki form of the perturbed
Klein-Gordon equation, as in equations~\eqref{KG1} and~\eqref{KG2}
in appendix~\ref{KG.alt}. }
 $h^2= 2a^2{\cal H}(1+q),$ and $h'\equiv\partial_N h.$ By differentiating this expression
 and using equations~\eqref{q.definition},~\eqref{2-q.itoU} and \eqref{H,q.prime}
one can express the coefficient $h'/h$ in~\eqref{DE.varphi.1}
in terms of the scalar field potential $V(\varphi_0)$, as follows:
\begin{equation} \label{partial.h}
\frac{h'}{h} = \frac{(2l V_{,\varphi} -V)}{2H^2},
\end{equation}
with $H$ given by~\eqref{H.ito.phi} in appendix~\ref{background}.

Equation~\eqref{DE.varphi.1} with~\eqref{partial.h} is the desired new form of the
perturbed Klein-Gordon equation at first order.
 By inspection it is clear that in the
super-horizon regime $l {}^{(1)}\!{\varphi}_{\mathrm c}\approx C$,
where $C$ is a spatial function, is a solution of this
equation as expected.
However, since~\eqref{DE.varphi.1} is a second order differential equation, it
will have two independent solutions, and the general solution
in the super-horizon regime can be written in the form
\begin{equation} \label{soln.SH1}
l {}^{(1)}\!{\varphi}_{\mathrm c}\approx C_1 +
C_2\int_{N_{init}}^N\frac{d{\bar N}}{h({\bar N})^2},
\end{equation}
which at first sight contradicts the earlier result~\eqref{phi_c.conserved} that
$l {}^{(1)}\!{\varphi}_{\mathrm c}\approx C$ \emph{is the general solution of
the perturbation equations in the super-horizon regime}. It follows that
the spatial function $C_2$ must be of order $O({\bf D}^2)$, making the second
term negligible in the super-horizon regime.

Although~\eqref{DE.varphi.1} can be solved explicitly
in the super-horizon regime as above
and also in the special case of power-law inflation
and in the slow-roll approximation\footnote
{This involves using conformal time  $\eta$ instead of $e$-fold
time $N$. See appendix~\ref{KG.alt} for the resulting alternative
forms of the Klein-Gordon equation.}
there is a restriction as regards
its overall applicability.
On recalling that $l=-1/(\varphi_0')$, it follows that
\emph{the coefficient~\eqref{partial.h} will be singular
whenever $\partial_N \varphi_0=0$}.\footnote
{Some implications of this singularity have been discussed by
Finelli and Brandenberger (1999)~\cite{finbra99}.}
 For example, during the
period of reheating that occurs at the end of inflation the scalar field
$\varphi_0$ oscillates, which means that $\partial_N \varphi_0$ will be zero
repeatedly. In order to avoid this singularity one can use
${}^{(1)}\!{\varphi}_{\mathrm c}$ as dependent variable.
The alternative form of the Klein-Gordon equation is given in
appendix~\ref{KG.alt}.

One can also use the above procedure to derive the perturbed
Klein-Gordon equation at second order. The leading order terms
in equations~\eqref{ucg_gov1} and~\eqref{infl.uni}
will be the same as at first order, but the equations will also have
source terms that are quadratic in the first order
perturbations $\phi_{\mathrm c},B_{\mathrm c}$ and $\varphi_{\mathrm c}$.\footnote{The
detailed expressions can be obtained from the equations in
UW2~\cite{uggwai18}. We do not give them because in this
paper we are just interested in the overall structure of the equations.}
At the first stage equations~\eqref{ucg1_varphi} will have the form
\begin{subequations}\label{ucg2_varphi}
\begin{align}
(1+q)\partial_N(l{}^{(2)}\!\hat{\varphi}_{\mathrm c}) &=
{\cal H}^{-2}{\bf D}^2({\cal H}{}^{(2)}\!{B}_{\mathrm c}) +
(1+q){\mathbb S}_{\varphi}, \label{ucg2_varphi.1} \\
\partial_N(a^2{}^{(2)}\!B_{\mathrm c} )&=
a^2 {\cal H}^ {-1}(1+q)l{}^{(2)}\!\hat{\varphi}_{\mathrm c}
+ {\mathbb S}_B,  \label{ucg2_varphi.2}
\end{align}
\end{subequations}
where we have chosen to use ${}^{(2)}\!\hat{\varphi}_{\mathrm c}$
instead of ${}^{(2)}\!{\varphi}_{\mathrm c}$ since this simplifies the source
term ${\mathbb S}_{\varphi}$ so that it has the property
\begin{equation} \label{S_varphi}
{\mathbb S}_{\varphi}={\cal O}({\bf D}^2).
\end{equation}
This can be confirmed by inspecting the various terms that contribute
to ${\mathbb S}_{\varphi}$. Here we note that we have used~\eqref{phi.varphi} to
replace ${}^{(1)}\!{\phi}_{\mathrm c}$ by ${}^{(1)}\!{\varphi}_{\mathrm c}$
in the source terms so that they depend only on the spatial derivatives
of ${}^{(1)}\!B_{\mathrm c}$ and ${}^{(1)}\!\varphi_{\mathrm c}$.
Eliminating ${}^{(2)}\!B_{\mathrm c}$ in equations~\eqref{ucg2_varphi}
as in the linear case leads to
\begin{equation} \label{DE.varphi.2}
\left(\partial_N^2+2\frac{h'}{h} \partial_N
- {\cal H}^{-2}{\bf D}^2 \right)(l {}^{(2)}\!\hat{\varphi}_{\mathrm c})=
\partial_N(h^2{\mathbb S}_{\varphi}) +  {\cal H}^{-2}{\bf D}^2 {\mathbb S}_B,
\end{equation}
where $h'/h$ is given by~\eqref{partial.h}.
Equation~\eqref{DE.varphi.2} is a new version of the
perturbed Klein-Gordon equation at second order.
We note that
there is one complication as regards the source term in~\eqref{DE.varphi.2}.
Since it depends on both  ${}^{(1)}\!\varphi_{\mathrm c}$ and ${}^{(1)}\!B_{\mathrm c}$
one has to express ${}^{(1)}\!B_{\mathrm c}$ in terms of ${}^{(1)}\!\varphi_{\mathrm c}$
using~\eqref{ucg1_varphi.1} in order to
make~\eqref{DE.varphi.2} a closed equation,\footnote{See for example,
Malik \emph{et al} 2008~\cite{maletal08} for an explicit
example of this process: the second order differential equation that arises from
the conservation of energy equation is equation (3.5) which becomes equation
(3.22) after the metric terms have been eliminated, in the slow-roll approximation.}
and this involves using the inverse Laplacian operator ${\bf D}^{-2}$.

An important property of equation~\eqref{DE.varphi.2} is
 that the total source
term on the right side of the equation is ${\cal O}({\bf D}^2)$
on account of~\eqref{S_varphi}, and this is due to using the hatted variable
${}^{(2)}\!\hat{\varphi}_{\mathrm c}$ instead of ${}^{(2)}\!{\varphi}_{\mathrm c}$.
Thus in the super-horizon regime, equation~\eqref{DE.varphi.2} reduces to the
DE
\begin{equation} \label{DE.varphi.2.S}
\left(\partial_N^2+2\frac{h'}{h} \partial_N
 \right)(l {}^{(2)}\!\hat{\varphi}_{\mathrm c})\approx 0,
\end{equation}
in complete analogy with equation~\eqref{DE.varphi.1},
which means that the general solution
has the same form as~\eqref{soln.SH1} with $l {}^{(2)}\!\hat{\varphi}_{\mathrm c}$
replacing $l {}^{(1)}\!{\varphi}_{\mathrm c}$.  Since equation~\eqref{phi_c.conserved}
shows that $l {}^{(2)}\!\hat{\varphi}_{\mathrm c}$
is a conserved quantity in general  the second term in the solution
must be negligible in the super-horizon regime,
as in the first order case.   Another similarity is that the
changes of variable made on the first order equation~\eqref{DE.varphi.1}
as in equations~\eqref{KG1}-\eqref{KG4},
which affect only the terms
on the left side of the equation,
can performed on equation~\eqref{DE.varphi.2} in an identical manner.
On the other hand changing from
${}^{(2)}\!\hat{\varphi}_{\mathrm c}$ to ${}^{(2)}\!{\varphi}_{\mathrm c}$
as the dependent variable using~\eqref{hat.varphi}
will complicate the equation~\eqref{DE.varphi.2} significantly by
 introducing additional  source terms that  are
 non-zero in the super-horizon regime.

Equations~\eqref{ucg2_varphi} or~\eqref{DE.varphi.2}, when transformed to Fourier
space both provide an algorithm for calculating
$l {}^{(2)}\!\hat{\varphi}_{\mathrm c}$ numerically.\footnote
{Numerical procedures for determining second order scalar perturbations
using the perturbed Klein-Gordon equation with ${}^{(2)}\!\varphi_{\mathrm c}$
rather than $l{}^{(2)}\!\varphi_{\mathrm c}$ as the dependent variable
have been given by Huston and Malik (2009)~\cite{husmal09} using the slow-roll
approximation, and by Huston and Malik (2011)~\cite{husmal11} in general.}
Using~\eqref{ucg2_varphi} would have
two advantages. First there is no need to eliminate ${}^{(1)}\!B_{\mathrm c}$ thereby
avoiding the introduction of the inverse Laplacian in the source terms, and second
the output directly determines two other quantities of
physical interest, namely the Bardeen potential $\psi_\mathrm{p}$ and the density perturbation
${\bdelta}_{\mathrm v}$, via the equations $\psi_\mathrm{p} = -{\cal H}B_{\mathrm c}$
and ${\bdelta}_{\mathrm v} = -\partial_N(l\varphi_{\mathrm c})$, generalized
to second order with source terms.

\section{Discussion}

In this paper we have considered second order perturbations of a flat Friedmann-Lema\^{i}tre
universe whose stress-energy content is a single minimally coupled scalar field. We have derived the general solution of the perturbed Einstein equations in
explicit form for this class of models when the perturbations are in the super-horizon
regime. We assumed an arbitrary potential and made
no use of the slow-roll approximation. We also showed that the Einstein equations in
the uniform curvature gauge lead to a new form of the perturbed Klein-Gordon
equation for linear perturbations using $l\varphi_{\mathrm c}$ as dependent  variable,
which we generalize to second order.

The perturbations of the scalar field have a simple form in the uniform curvature gauge,
which reflects the fact that the perturbations admit a conserved
quantity. Although second order perturbations of scalar fields
minimally coupled to gravity
have been studied extensively during the past fifteen years
in connection with inflation, starting with
Acquaviva \emph {et al} (2003)~\cite{acqetal03} and
Maldacena (2003)~\cite{mal03} (see also, for example,
Finelli \emph {et al}~\cite{finetal04,finetal06},
Malik (2005)~\cite{mal05} and Vernizzi (2005)~\cite{ver05}), the conserved
quantity $l{}^{(2)}\!\hat{\varphi}_{\mathrm c}$ given by~\eqref{hat.varphi.new}
and the general solution for ${}^{(2)}\!\varphi_{\mathrm c}$
in equation~\eqref{varphi_c.solution} have
not been given previously, to the best of our knowledge. However, we note that
Finelli \emph{et al} (2004)~\cite{finetal04} have derived an approximate
solution for ${}^{(2)}\!\varphi_{\mathrm c}$ using the perturbed
Klein-Gordon equation at second order (their equation (28)). They
impose the slow-roll approximation and use
the potential $V=\sfrac12 m^2 \varphi_0^2$
that corresponds to chaotic inflation. We have not
been able to relate their solution which is described in equations (53)-(57)
in~\cite{finetal04} to our general result. In addition, some of the change of
variable formulas that we use in section~\ref{cons.quantity} have been given
previously but in a more complicated form. For example
equation (8) in Finelli \emph{et al} (2006)~\cite{finetal06}
corresponds to our equation~\eqref{psi-varphi2},
while equations (3.3)-(3.4)
in Maldacena (2003)~\cite{mal03} correspond to the special case of
our equation~\eqref{psi-varphi2}
when the gauge on the right side is chosen to be the uniform curvature gauge.

\begin{appendix}

\section{Cosmological scalar fields as perfect fluids~\label{scalar.field}}

We are considering a flat FL universe in which the matter-energy
content is a single minimally coupled scalar field $\varphi$.
The stress-energy tensor is of the form
\begin{subequations}
\begin{equation}\label{sfT_ab}
T^a\!_b = \bna^a\varphi\bna\!_b\varphi - \left[\sfrac12 \bna^c\varphi\bna\!_c\varphi +
V(\varphi)\right]\delta^a\!_b ,
\end{equation}
and the conservation equation leads to the Klein-Gordon equation
$\bna^c\bna\!_c\varphi - V\!_{,\varphi} = 0$,
where the potential $V(\varphi)$ has to be specified. In cosmology this stress-energy tensor has the perfect fluid form
\begin{equation}\label{pf}
T^a\!_b = \left(\rho + p\right)\!u^a u_b + p\delta^a\!_b,
\end{equation}
with
\begin{equation}\label{T_ab,scalar}
\rho + p = -  \bna^a\varphi \bna\!_a\varphi, \qquad \rho - p = 2V(\varphi),
\qquad u_a = \frac{\bna\!_a\varphi}{\sqrt{- \bna^a\varphi \bna\!_a\varphi}} .
\end{equation}
\end{subequations}
%

\subsection{Background equations \label{background}}

In a spatially flat background the Friedmann
equation\footnote{We use units with $c=\hbar=1$ and
$8\pi G=1/(M_{Pl})^2 = 1$, where $M_{Pl}$  is the reduced Planck mass
and $G$ the gravitational constant.}
and the conservation of energy equation read\footnote
{We remind the reader that a $'$ denotes the derivative with respect to $N$.}
\begin{subequations} \label{basic.relations}
\begin{equation} \label{prop.of.rho}
3H^2 = \rho_0, \qquad \rho_0' = -3(\rho_0+p_0),
\end{equation}
where $H$ is the background Hubble variable while $\rho_0$
and $p_0$ are the background
energy density and pressure, respectively.
We introduce the standard matter variables
$w=p_0/\rho_0$ and $c_s^2=p_0'/\rho_0'$.
Using these equations and the definition $1+q=- H'/H$
it follows that
\begin{equation} \label{prop.of.w}
3(1+w)=2(1+q), \qquad  w'=3(1+w)(w-c_s^2).
\end{equation}
When evaluated on the FL background, equation~\eqref{T_ab,scalar} leads to
\begin{equation}\label{rho,p for phi}
\rho_0 + p_0 = H^2(\varphi_0')^2, \qquad  \rho_0 - p_0 = 2V(\varphi_0).
\end{equation}
\end{subequations}
We can use equations~\eqref{basic.relations} to express $H^2$,
$w$ and $c_s^2$ in terms of $\varphi_0$ and $V(\varphi_0)$. To indicate that $w$
and $c_s^2$ describe a scalar
field we will label them as $w_{\varphi}$ and $c_{\varphi}^2$.
The resulting expressions are as follows:
\begin{subequations} \label{wcs_N}
\begin{align}
H^2&=\frac{2V(\varphi_0)}{6-(\varphi_0')^2},  \label{H.ito.phi} \\
w_{\varphi} &= -1 +\sfrac13 (\varphi_0')^2, \label{w.ito.phi} \\
c_{\varphi}^2 &= 1 + \frac{(6-(\varphi_0')^2)}{3\,\varphi_0'}
\frac{V_{,\varphi}}{V},
\end{align}
\end{subequations}
where $V_{,\varphi}$ is the derivative of
$V(\varphi_0)$ with respect to $\varphi_0$.
It follows from~\eqref{prop.of.w} and~\eqref{wcs_N} that
%
\begin{equation} \label{2-q.itoU}
2-q=\frac{V}{H^2}, \qquad 1-c_{\varphi}^2=
-\frac{2}{3\varphi_0'}\frac{V_{,\varphi}}{H^2}.
\end{equation}
Further, one can derive the background Klein-Gordon equation~\eqref{KG_N}
by differentiating~\eqref{H.ito.phi}, and then use the result to obtain
\begin{equation} \label{w-c^2}
w_{\varphi}-c_{\varphi}^2 =\frac23\left(\frac{\varphi_0''}{\varphi_0'}\right).
\end{equation}
We will also need the following result
\begin{equation} \label{H,q.prime}
 q'= 3(1+q)(w_{\varphi}-c_{\varphi}^2),
\end{equation}
which is an immediate consequence of~\eqref{prop.of.w}.

We end this section with a brief digression on the Hubble flow
functions $\varepsilon_n, \, n=1,2,3\dots$ which define the
slow-roll regime,
although we do not use this approximation. These
functions are defined by
$\varepsilon_1=-H'/H,\, \varepsilon_{n+1}=\varepsilon_n'/\varepsilon_n,\,
n=1,2,3\dots$ (see for example,
Martin (2016)~\cite{mar16}, equation (5)). It follows
from $1+q=- H'/H$ and equations~\eqref{prop.of.w},~\eqref{w.ito.phi}
and~\eqref{w-c^2} that the Hubble flow functions
are related to the scalar field  according to
\begin{equation}
\varepsilon_1= 1+q = \sfrac12(\varphi_0')^2,
\qquad \varepsilon_2=\frac{q'}{1+q}=2\left(\frac{\varphi_0''}{\varphi_0'}\right).
\end{equation}
%

\subsection{Perturbations of the scalar field \label{app:scalar} }

For a scalar field we can express the matter variables
$({}^{(r)}\!{\bdelta}, {}^{(r)}\!{P}, {}^{(r)}\!{V})$,
where ${}^{(r)}\!{P}={}^{(r)}\!{p}/(\rho_0+p_0)$,
in terms of the scalar field perturbations ${}^{(r)}\!{\varphi}$ and the metric variables
$\phi$ and $B$ using equations~\eqref{sfT_ab} and~\eqref{T_ab,scalar}.
The results at first order are
\begin{subequations} \label{scalar1}
\begin{align}
{}^{(1)}\!{\bdelta} + {}^{(1)}\!{P} &=
 -2(l{\partial}_{N}{}^{(1)}\!{\varphi} +{}^{(1)}\!\phi) ,\label{scalar1.1} \\
{}^{(1)}\!{\bdelta} - {}^{(1)}\!{P} &=
2H^{-2}\, lV_{,\varphi} \,l{}^{(1)}\!{\varphi} ,  \label{scalar1.2}   \\
{\cal H}{}^{(1)}\!{V} &= l{}^{(1)}\!{\varphi}, \label{scalar1.3}
\end{align}
\end{subequations}
where $l=-(\varphi_0')^{-1}$
is given by~\eqref{def_lambda}.
The results at second order are:\footnote{Formulas
for ${}^{(r)}\!T^a\!_b$, $r=1,2$, for the stress-energy tensor~\eqref{sfT_ab} have been given
for example, by Acquaviva \emph{et al} (2003)~\cite{acqetal03}
(see equations (9)-(16)), by Malik (2007)~\cite{mal07}
(see equations (C12)-(C14)) and by Nakamura (2009)~\cite{nak09a}
(see equations (4.45)-(4.52)).
The expressions we have given can be obtained using the relation
${}^{(2)}\!T^a\!_b = (\rho_0 +p_0)({\mathsf T}^a\!_b + {\mathbb T}^a\!_b)$, see UW2~\cite{uggwai18}.}
\begin{subequations} \label{scalar2}
\begin{align}
{}^{(2)}\!{\bdelta} + {}^{(2)}\!{P} &=
-2(l{\partial}_{N}{{}^{(2)}\!\varphi} + {}^{(2)}\!\phi) +
2\left(2{}^{(1)}\!\phi +l{\partial}_{N}{{}^{(1)}\!\varphi}\right)^2 +
2{\cal H}^{-2}\left({\bf D}(l{{}^{(1)}\!\varphi} - {\cal H} {}^{(1)}\!B)\right)^2 , \label{scalar2.1} \\
{}^{(2)}\!{\bdelta} - {}^{(2)}\!{P} &=
2H^{-2}\left(lV_{,\varphi}\,l{}^{(2)}\!{\varphi} +
V_{,\varphi\varphi} (l{{}^{(1)}\!\varphi})^2\right),  \\
{\cal H}{}^{(2)}\!{V} &=l {}^{(2)}\!{\varphi} -
{\cal S}^i\left[({{}^{(1)}\!\bdelta} + {{}^{(1)}\!P}){\bf D}_i {\cal H}{{}^{(1)}\!V}\right]. \label{scalar2.3}
\end{align}
\end{subequations}
Before continuing we note two properties of the perturbations
of the scalar field that are obtained immediately  by choosing the
total matter gauge in~\eqref{scalar1} and~\eqref{scalar2}, namely that
\begin{equation}  \label{sf.prop1}
{}^{(r)}\!\varphi_{\mathrm v} = 0,\qquad r=1,2.
\end{equation}
and hence that
\begin{equation}  \label{sf.prop2}
{}^{(r)}\!{P}_{\mathrm v} = {}^{(r)}\!{\bdelta}_{\mathrm v},\qquad r=1,2.
\end{equation}

We now derive expressions for the non-adiabatic
pressure perturbations ${}^{(r)}\!\Gamma$, $r=1,2$ for
a perturbed scalar field. The general expressions are given
in UW2~\cite{uggwai18} (see equation (23)), which we repeat here
\begin{subequations} \label{gamma_general}
\begin{align}
{}^{(1)}\!{\Gamma} &= {}^{(1)}\!{P} - c_s^2\, {}^{(1)}\!{\bdelta},\label{gamma_1} \\
{}^{(2)}\!{\Gamma} &= {}^{(2)}\!{P} - c_s^2\,{}^{(2)}\!{\bdelta} +
\sfrac13 (\partial_N c_s^2)({{}^{(1)}\!\bdelta})^2 +
\sfrac23{{}^{(1)}\!\bdelta}\left[\partial_N -
3(1 + c_s^2)\right]{{}^{(1)}\!\Gamma}. \label{gamma_2}
\end{align}
\end{subequations}
The expressions on the right side are independent of choice of timelike gauge
once the spatial gauge has been fixed as in UW2~\cite{uggwai18}.
In the present situation we evaluate them in the total matter gauge
and use~\eqref{sf.prop2} which leads to
\begin{subequations} \label{gamma_sf}
\begin{align}
{}^{(1)}\!{\Gamma} &= (1-c_s^2){}^{(1)}\!{\bdelta}_{\mathrm v}, \label{gamma_sf1} \\
{}^{(2)}\!{\Gamma} &= (1-c_s^2){}^{(2)}\!{\bdelta}_{\mathrm v} +
\sfrac13 (\partial_N c_s^2)({{}^{(1)}\!\bdelta}_{\mathrm v})^2 +
\sfrac23 {{}^{(1)}\!\bdelta_{\mathrm v}}\left(\partial_N - 3(1+c_s^2) \right)\!{{}^{(1)}\!\Gamma}. \label{gamma_sf_2}
\end{align}
\end{subequations}
The constraints~\eqref{scalar1.3} and~\eqref{scalar2.3} play
a central role in that they determine the
perturbations of the scalar field in terms of the velocity perturbations
in an arbitrary gauge. We can write~\eqref{scalar2.3} in terms of
the hatted variables, in a form that will be useful later, as follows.
We combine~\eqref{scalar1.1} and~\eqref{scalar1.3} to obtain
\begin{equation}  \label{delta+P}
{}^{(1)}\!{\bdelta} + {}^{(1)}\!{P} =
 -2\left(({\partial}_{N}+1+q)({\cal H}{}^{(1)}\!V) +{}^{(1)}\!\phi
 - \sfrac32(1+c_{\varphi}^2){\cal H}{}^{(1)}\!V\right),
\end{equation}
using $l'/l=-(1+q)+\sfrac32(1+c_{\varphi}^2)$.
The perturbed conservation of momentum equation
(UW2~\cite{uggwai18}, section 4) when
applied to a scalar field using~\eqref{gamma_sf1} gives
\begin{equation}
({\partial}_{N} +1+q)({\cal H}{}^{(1)}\!V) + {}^{(1)}\!\phi = -{}^{(1)}\!{\bdelta}_{\mathrm v},
\end{equation}
which when substituted in~\eqref{delta+P} yields
\begin{equation}
{}^{(1)}\!{\bdelta} + {}^{(1)}\!{P} =3(1+c_{\varphi}^2){\cal H}{}^{(1)}\!V + 2{}^{(1)}\!{\bdelta}_{\mathrm v}.
\end{equation}
We substitute this expression in~\eqref{scalar2.3} and introduce
the hatted variables ${}^{(2)}\!{\hat{\varphi}}$ defined in~\eqref{hat.varphi} and
${\cal H}{}^{(2)}\!{\hat{V}} =
{\cal H}{}^{(2)}\!{V} + (1+q)({\cal H} {}^{(1)}\!{V})^2.$
Together with~\eqref{scalar1.3} we obtain
\begin{subequations} \label{varphi_V}
\begin{align}
l{}^{(1)}\!{\varphi}&={\cal H}{}^{(1)}\!{V} ,    \label{scalar1.3a}  \\
l {}^{(2)}\!{\hat{\varphi}} &= {\cal H}{}^{(2)}\!{\hat V} +
2{\cal S}^i[{\bdelta}_{\mathrm v}{\bf D}_i {\cal H}{V}].   \label{scalar2.3a}
\end{align}
\end{subequations}
%

\subsection{Alternative forms for the perturbed Klein-Gordon equation \label{KG.alt}}

We first  transform the differential equation~\eqref{DE.varphi.1}
to Fourier space (${\bf D}^2\rightarrow -k^2$) and introduce
conformal time $\eta$ obtaining
\begin{equation} \label{KG1}
\left(\partial_{\eta}^2\,+
2({\partial_{\eta}z }/{z}) \partial_{\eta}
+k^2\right)(l {}^{(1)}\!{\varphi}_{\mathrm c})=0, \qquad z=h/\sqrt{\cal H}=a/l.
\end{equation}
We make the transition from $N$ to $\eta$ by using
\begin{equation}
\partial_{\eta} = {\cal H}\partial_{N}, \qquad
\partial_{\eta}^2 = {\cal H}^2(\partial_{N}^2-q\partial_{N}).
\end{equation}
Alternatively one can transform the above differential equation
to the so-called Mukhanov-Sasaki form by
scaling $l {}^{(1)}\!{\varphi}_{\mathrm c}$ with $z$:
\begin{equation} \label{KG2}
\left(\partial_{\eta}^2\,-
({\partial_{\eta}^2z }/{z})
-k^2)\right)(a {}^{(1)}\!{\varphi}_{\mathrm c})=0, \qquad z=a/l.
\end{equation}
On recalling that the comoving curvature perturbation is given by
${\cal R}=\psi_{\mathrm v}=-l{}^{(1)}\!{\varphi}_{\mathrm c}$
these differential equations can be written with
${\cal R}$ and $z{\cal R}$, respectively,  as the dependent variable.
See, for example, Weinberg (2008)~\cite{wei08}, equation (10.3.1) and page 481,
and Durrer (2008)~\cite{dur08}, equation (3.35) and page 113, respectively.
In the case of power-law inflation and when using the slow-roll approximation
equations~\eqref{KG1} and~\eqref{KG2} reduce to particular forms of Bessel's equation,
and hence can be solved. See for example~\cite{dur08},
pages 113-115, and in~\cite{wei08}, pages 481-482 and 488-491.

Finally, if we use ${\varphi}_{\mathrm c}$ as the dependent variable,
the differential equation~\eqref{DE.varphi.1} assumes the following form:
\begin{equation} \label{KG3}
 \partial_N^2{\varphi}_{\mathrm c} +
 \frac{V}{H^2} \partial_N {\varphi}_{\mathrm c} +
 \frac{(V_{,\varphi\varphi}+2\varphi_0' V_{,\varphi} +
 (\varphi_0' )^2V)}{H^2}{\varphi}_{\mathrm c} -
{\cal H}^{-2}\,{\bf D}^2{\varphi}_{\mathrm c}=0,
\end{equation}
(see  for example,
Huston and Malik (2009)~\cite{husmal09}, equation (3.9), in the
Fourier domain, noting that $\delta \dot{\varphi}$ denotes differentiation with
respect to $N$ in this reference.). If we change to conformal time we obtain
\begin{equation}  \label{KG4}
\partial_{\eta}^2{\varphi}_{\mathrm c} +2{\cal H} \partial_{\eta}{\varphi}_{\mathrm c} +
a^2\left(V_{,\varphi\varphi}+2{\varphi_0'} V_{,\varphi} +
(\varphi_0' )^2V\right){\varphi}_{\mathrm c} -
{\bf D}^2{\varphi}_{\mathrm c}=0,
\end{equation}
with $\varphi_0'=\partial_\eta \varphi_0/{\cal H}$ (see for
example~\cite{husmal09}, equation (2.13), noting that $'$
denotes differentiation with respect to conformal time in this reference.)
One can see by inspection that the coefficients of~\eqref{KG3}
and~\eqref{KG4} are well-defined when $\varphi_0'=0.$

\end{appendix}

\bibliographystyle{plain}
\bibliography{../Bibtex/cos_pert_papers}

\end{document}